\documentclass[global,referee]{svjour}
\usepackage{graphicx}
\usepackage{epsfig}
\usepackage{dcolumn}
\usepackage{natbib}
\usepackage{amssymb}

\journalname{Int. Journal of Fracture}

\def\dj{\hbox{d\kern-0,347em \vrule width0,3em height1,252ex
depth-1,21ex \kern0,051em}}

\begin{document}

\title{Fracture size effects from disordered lattice models}
\author{Mikko J. Alava\inst{1}, Phani K.V.V. Nukala\inst{2}, Stefano Zapperi
\inst{3} \inst{4}} \institute{Department of Engineering Physicsry of
Physics, Helsinki University of Technology, FIN-02015 HUT, Finland
\and Computer Science and Mathematics Division, Oak Ridge National
Laboratory, Oak Ridge, TN~37831-6164, USA \and CNR-INFM, S3,
Dipartimento di Fisica, Universit\`a di Modena e Reggio Emilia, Via
G. Campi 213A,  Modena, Italy \and ISI Foundation, Viale S. Severo
65, 10133 Torino, Italy}
\date{Received: date / Revised version: date}
\maketitle
\begin{abstract}
We study size effects in the fracture strength of notched disordered
samples using numerical simulations of lattice models for fracture.
In particular, we consider the random fuse model, the random spring
model and the random beam model, which all give similar results.
These allow us to establish and understand the crossover between a
regime controlled by disorder-induced statistical effects  and a
stress-concentration controlled regime ruled by fracture mechanics.
The crossover is described by a scaling law that accounts for the
presence of fracture process zone which we quantify by
averaging over several disordered configurations of the model. The models
allow to study the development of the fracture process zone as the
load is increased and to express this in terms of
crack resistance (R-curve). 
\end{abstract}

\section{Introduction}

Understanding how materials break is a fundamental open problem of
science and engineering. The difficulties stem from the non-trivial
dependence of the fracture strength on the characteristic
lengthscales of the samples, as already noted by Leonardo da Vinci,
who measured the carrying-capacity of metal wires of varying length
\cite{leonardo40}. He observed that the longer the wire, the less
weight it could sustain. The reason for this behavior is rooted in
the structural disorder present in the material: the strength is
dominated by the weakest part (subvolume) of the sample and its
distribution can in principle be computed using extreme value
statistics \cite{gumbel}. Longer wires have more weak parts and are
thus bound to fail at smaller loads on average. The quantitative
understanding of this statistical size effect is difficult, since
the important low-strength tails of the strength probability
distribution are not easy to sample and since the material
properties are often history-dependent. For instance, in
quasi-brittle materials such as concrete and many other composites,
where sample failure is preceded by significant damage accumulation
\cite{vanmierbook}.

An important engineering scenario and a typical experimental setting
is to study the size effect in the presence of a pre-existing flaw,
{\it a notch}. Failure in this case is determined by the competition
between deterministic effects, due to the stress enhancement created
by notch, and the response of the disordered material around the defect to the
stress concentration \cite{bazantbook,bazant99}. This includes the
stochastic damage accumulation. The effect of disorder can be
treated in an effective medium sense by defining a Fracture Process
Zone (FPZ) around the crack tip.
For quasi-brittle materials, the size of this FPZ may not be
negligible compared to the system size.
Conversely, for small notches failure is more influenced by the FPZ
than the notch itself, and depends on statistical disorder effects.
Experimentally, it has been demonstrated that the critical crack may
nucleate and propagate far from the pre-existing notch \cite{rosti01}
in that case.

The existing theories on the size effect start from linear elastic
fracture mechanics (LEFM). Several formulations have been proposed
in the literature and partly compared with experiments
\cite{bazant99,hu92,karihaloo99,bazant00,bazant04b,morel00,morel02}.
The problem is how to extend LEFM in the presence of disorder and concomitant damage.

In LEFM the Griffith's stability criterion or equation for the
maximum stress the specimen can bear reads $\sigma_c \sim
K_c/\sqrt{a_0}$. Here $a_0$ is the linear size of the crack and the
critical stress intensity factor $K_c \sim \sqrt{E G_c}$ is a
function of the fracture toughness $G_c$ and the elastic modulus $E$
\cite{griffith20}. In the size-effect law proposed by Bazant for
quasi-brittle materials \cite{bazant99,bazant00,bazant04b}, the
Griffith expression is generalized considering an additional
lengthscale $\xi$  due to the presence of a FPZ
\begin{equation}
\sigma_c = K_c/\sqrt{\xi + a_0}.
\label{baz}
\end{equation}
Equation~(\ref{baz}) takes into account the both limits of a very
large notch compared to the FPZ size and that of a very small notch.
In the limit $\xi/a_0 \ll 0$ one has an expression that follows the
LEFM scaling, $\sigma_c \sim 1/\sqrt{a_0}$. In the opposite limit of
$a_0\rightarrow0$, the average strength is taken to be constant.
Eq.~(\ref{baz}) has been shown to be in agreement with several
experimental data sets \cite{bazant04b}. However, three fundamental
questions can be asked: first, does Eq. (\ref{baz}) incorporate all
the important effects? Second, what is the fracture toughness $G_c$
in the presence of disorder? Third, where does the FPZ scale $\xi$
originate and how does it depend again on the disorder?

In this work we clarify the role of the disorder in the failure of
notched quasi-brittle specimens using extensive simulations of
disordered lattice models for fracture \cite{alava06}. A brief
account of the results for the RFM has been published in
Ref.~\cite{alava08}. In more detail, we consider the random fuse
model (RFM), the random spring model (RSM) and the random beam model
(RBM). We consider the failure of notched disordered samples and
provide a microscopic justification of Eq. (\ref{baz}). Studying the
size scaling of strength by  extensive numerical simulations is a
difficult task due to the different length scales involved and the
need of significant statistical averaging. We vary the disorder,
which we model as a locally varying random failure threshold, and
show that it plays a crucial role in determining the size effect. In
particular, a lengthscale $\xi$ emerges from the simulations and can
be shown to be directly related to the FPZ size. Finally, for notch
sizes smaller than a critical length $a_c$, we observe a cross-over
to the inherent, sample-size dependent strength of the unnotched
sample. We outline the scaling theory to account for these effects.
We show that the RFM results are confirmed in the RSM and the RBM.
We also study the growth of the FPZ, showing that it is independent
of the notch size $a_0$ and the system size $L$.

\section{Models}

\subsection{Random fuse model}

In the RFM \cite{deArcangelis85}, we consider a
triangular lattice of linear size $L$ with a central notch of length
$a_0$. Each fuse has the same conductance and a
random breaking threshold $t$. This represents a locally varying fracture
toughness/strength. The $t$ lie between 0 and 1, with a cumulative 
distribution $P(t) = t^{1/D}$, where $D$ represents a
quantitative measure of disorder. The larger $D$ is, the stronger the
disorder. The burning of a fuse occurs irreversibly, whenever the electrical
current in the fuse exceeds breaking threshold $t$ of the
fuse. Periodic boundary conditions are imposed in the horizontal
direction to simulate an infinite system and a constant voltage
difference, $V$, is applied between the top and the bottom of lattice
system bus bars. Numerically, a unit voltage difference, $V = 1$, is
set between the bus bars and the Kirchhoff equations are solved to
determine the current flowing in each of the fuses. Subsequently, for
each fuse $j$, the ratio between the current $i_j$ and the breaking
threshold $t_j$ is evaluated, and the bond $j_c$ having the largest
value, $\mbox{max}_j \frac{i_j}{t_j}$, is irreversibly removed
(burnt).  The current is redistributed instantaneously after a fuse is
burnt implying that the current relaxation in the lattice system is
much faster than the breaking of a fuse.  Each time a fuse is burnt,
it is necessary to re-calculate the current redistribution in the
lattice to determine the subsequent breaking of a bond.  The process
of breaking of a bond, one at a time, is repeated until the lattice
system fails completely. In the present simulations, we have considered
various notch sizes for $L=64,128,192,256,320$ and $D=0.1,0.3,0.5,0.6,0.75$.

\subsection{Random spring model}

In the RSM, we consider a triangular lattice with nodes connected by linear springs
with unit stiffness \cite{sahimi86,hansen89C,sahimi931,sahimi932,nukala05}.
As for the RFM, the bond breaking
thresholds, $t$, are randomly distributed based on a thresholds cumulative 
probability distribution, $P(t)=t^{1/D}$ for $t \in [0,1]$.
The bond breaks irreversibly, whenever the force in the spring exceeds the
breaking threshold force value, $t$, of the spring. Periodic boundary
conditions are imposed in the horizontal direction and a displacement
difference is applied between the top and the bottom of the lattice system.
Numerically, a unit displacement, $\Delta = 1$, is applied at the top
of the lattice system and the equilibrium equations are solved to
determine the force in each of the springs. Subsequently, for each
bond $j$, the ratio between the force $f_j$ and the breaking threshold
$t_j$ is evaluated, and the bond $j_c$ having the largest value,
$\mbox{max}_j \frac{f_j}{t_j}$, is irreversibly removed. The forces
are redistributed instantaneously after a bond is broken implying that
the stress relaxation in the lattice system is much faster than the
breaking of a bond. Each time a bond is broken, it is necessary to
re-equilibrate the lattice system in order to determine the subsequent
breaking of a bond.  The process of breaking of a bond, one at a time,
is repeated until the lattice system falls apart. For the RSM, we
consider a triangular lattice network of size $L$ with a notch
of size $a_0$. In the present simulations, we have considered
various notch sizes for $L=128,256$ and $D=0.5,0.6$.

\subsection{Random beam model}
In the random thresholds beam model (RBM) \cite{roux85,herrmann89},
we consider a two-dimensional triangle
lattice system of linear size $L$. The vectorial RBM has three
degrees of freedom (x-translation $u$, y-translation $v$, and a rotation $\theta$ about z axis) at each of
the lattice nodes (sites), and each of the bonds (beams) in the lattice connects two nearest neighbor
nodes. We assume that the beams are connected rigidly at each of the nodes such that the
angle between any two beams connected at a node remains unaltered during the deformation
process. These nodal displacements and rotations introduce conjugate forces and bending moments
in the beam members. Using Timoshenko beam theory \cite{przemieniecki}, which includes shear deformations of the beam
cross-section in addition to the usual axial deformation of cross-sections, the {\it local} stiffness matrix
for a beam element that relates the local nodal displacements and
rotations to local nodal forces and bending moments in the beam's {\it local} coordinate system is
given by
\begin{eqnarray}
{\bf K}_{local} & = & \left[\begin{array}{ccccccccc}
\frac{EA}{\ell_b} & 0 & 0 & -\frac{EA}{\ell_b} & 0 & 0 \\
 & \frac{12EI}{(1+\alpha) \ell_b^3} & \frac{6EI}{(1+\alpha) \ell_b^2} & 0 & -\frac{12EI}{(1+\alpha) \ell_b^3} & \frac{6EI}{(1+\alpha) \ell_b^2} \\
 & & \frac{(4+\alpha) EI}{(1+\alpha) \ell_b} & 0 & -\frac{6EI}{(1+\alpha) \ell_b^2} & \frac{(2-\alpha) EI}{(1+\alpha) \ell_b} \\
 & & & \frac{EA}{\ell_b} & 0 & 0 \\
 & SYM & & & \frac{12EI}{(1+\alpha) \ell_b^3} & -\frac{6EI}{(1+\alpha) \ell_b^2} \\
 & & & & & \frac{(4+\alpha) EI}{(1+\alpha) \ell_b} \\
\end{array} \right] \label{Klocal}
\end{eqnarray}
where $E$ is the Young's modulus, $G$ is the shear modulus, $A$ is
the beam cross-sectional area, $I$ is the moment of inertia of beam
cross-section, $\ell_b$ is the length of the beam, and $\alpha =
\frac{12EI}{GA \ell_b^2}$ is the shear correction factor, which
denotes the ratio of bending stiffness to the shear stiffness. If
shear deformation of beam cross-section is negligible, then $\alpha
= 0$ and the Timoshenko beam theory reduces to Euler-Bernoulli beam
theory. Equation \ref{Klocal} presents a relation between local
nodal displacements and rotations ${\bf d}_{\ell} = (u_{li}, v_{li},
\theta_{li}, u_{lj}, v_{lj}, \theta_{lj})^T$ and local forces and
moments ${\bf F}_{\ell} = (F_{li}, V_{li}, M_{li}, F_{lj}, V_{lj},
M_{lj})^T$. In this setting, the subscript $l$ refers to local
quantities, the superscript $T$ represents transpose of a vector or
a matrix, $i$ and $j$ refer to $i$-th and $j$-th nodes of the beam,
and $F$, $V$, and $M$ refer to axial force, shear force, and bending
moments respectively.

Equilibration of the lattice system is achieved by first transforming these local
quantities (${\bf d}_{\ell}$ and ${\bf F}_{\ell}$) into global quantities
${\bf d} = (u_{i}, v_{i}, \theta_{i}, u_{j}, v_{j}, \theta_{j})^T$ and
${\bf F} = (F_{i}, V_{i}, M_{i}, F_{j}, V_{j}, M_{j})^T$
through a coordinate transformation ${\bf T}$ such that ${\bf d}_{\ell} = {\bf T} {\bf d}$,
${\bf F}_{\ell} = {\bf T} {\bf F}$, and ${\bf K} = {\bf T}^T {\bf K}_{local} {\bf T}$, and then satisfying
equilibrium equations at each node such that
\begin{eqnarray}
\Sigma_{<ij>} F_x & = & 0; ~~~ \Sigma_{<ij>} F_y = 0; ~~~ \Sigma_{<ij>} M = 0 \label{eqbm}
\end{eqnarray}
where $\Sigma_{<ij>}$ implies that the summation is carried over all the
intact bonds $<ij>$ joining at node $i$. In the above discussion, the transformation
matrix ${\bf T}$ is given by
\begin{eqnarray}
{\bf T} & = & \left[\begin{array}{cccccc}
{\bf Q} & {\bf 0} \\
{\bf 0} & {\bf Q} \\
\end{array} \right] \label{Tarray}
\end{eqnarray}
where
\begin{eqnarray}
{\bf Q} & = & \left[\begin{array}{ccccc}
c & s & 0 \\
-s & c & 0 \\
0 & 0 & 1 \\
\end{array} \right] \label{Qarray}
\end{eqnarray}
and $c = \cos(\beta)$, $s = \sin(\beta)$ refer to the direction cosines of the beam with $\beta$
representing the angle between the beam axis and the $x$-direction.

In the present simulation, we start with a notched lattice system with
beams having unit length, unit square cross-section and
Young's modulus $E = 1$. This results in a unit axial stiffness ($EA/\ell_b = 1$) and
bending stiffness ($12 EI/\ell_b^3 = 1$) for each of the beams in the lattice system.
Since the beam can deform in two independent deformation modes (axial and bending), we
assume randomly distributed bond breaking axial
and bending thresholds, $t_a$ and $t_b$, based on thresholds cumulative 
probability distributions, $P_a(t_a)$ and $P_b(t_b)$ respectively. As
in the other models, the cumulative distributions are defined as $P(t)= t^{1/D}$ in $[0,1]$.

The failure criterion for a beam is defined through an axial force and bending moment
interaction equation (similar to von-Mises criterion in metal plasticity) given by
\begin{eqnarray}
r & \equiv & \left(\frac{F}{t_a}\right)^2 + \frac{\mbox{max}(|M_i|, |M_j|)}{t_b} = 1 \label{inter}
\end{eqnarray}
The beam breaks irreversibly, whenever the failure criterion $r \ge 1$. Periodic boundary
conditions are imposed in the horizontal direction and a constant unit displacement
difference is applied between the top and the bottom of lattice system.

Numerically, a unit displacement, $\Delta = 1$, is applied at the top
of the lattice system and the equilibrium equations (Eq. \ref{eqbm}) are solved to
determine the force in each of the springs. Solution of Eq. \ref{eqbm}
results in global displacements and rotations ${\bf d}$, using which
the local displacements ${\bf d}_{\ell} = {\bf T} {\bf d}$ and the
local forces ${\bf F}_{\ell} = {\bf K}_{local} {\bf d}_{\ell}$ are computed
for each of the intact beams. Subsequently, for each
bond $k$ with nodes $i$ and $j$, the quantities $a_{k} = \left(\frac{F}{t_a}\right)^2$ and
$b_k = \frac{\mbox{max}(|M_i|, |M_j|)}{t_b}$ are evaluated,
and the bond $k_c$ having the smallest value,
\begin{eqnarray}
r_k & = & \frac{-b_k + \sqrt{b_k^2 + 4 a_k}}{2 a_k} \label{req}
\end{eqnarray}
is irreversibly removed (When $a_k = 0$, then $r_k =
\frac{1}{b_k}$). The forces are redistributed instantaneously after
a bond is broken implying as in the other models that the stress
relaxation in the lattice system is much faster than the breaking of
a bond. Each time a bond is broken, it is necessary to
re-equilibrate the lattice system in order to determine the
subsequent breaking of a bond.  The process of breaking of a bond,
one at a time, is repeated until the lattice system falls apart.

\begin{figure}
\begin{center}
\includegraphics[height=3.5cm]{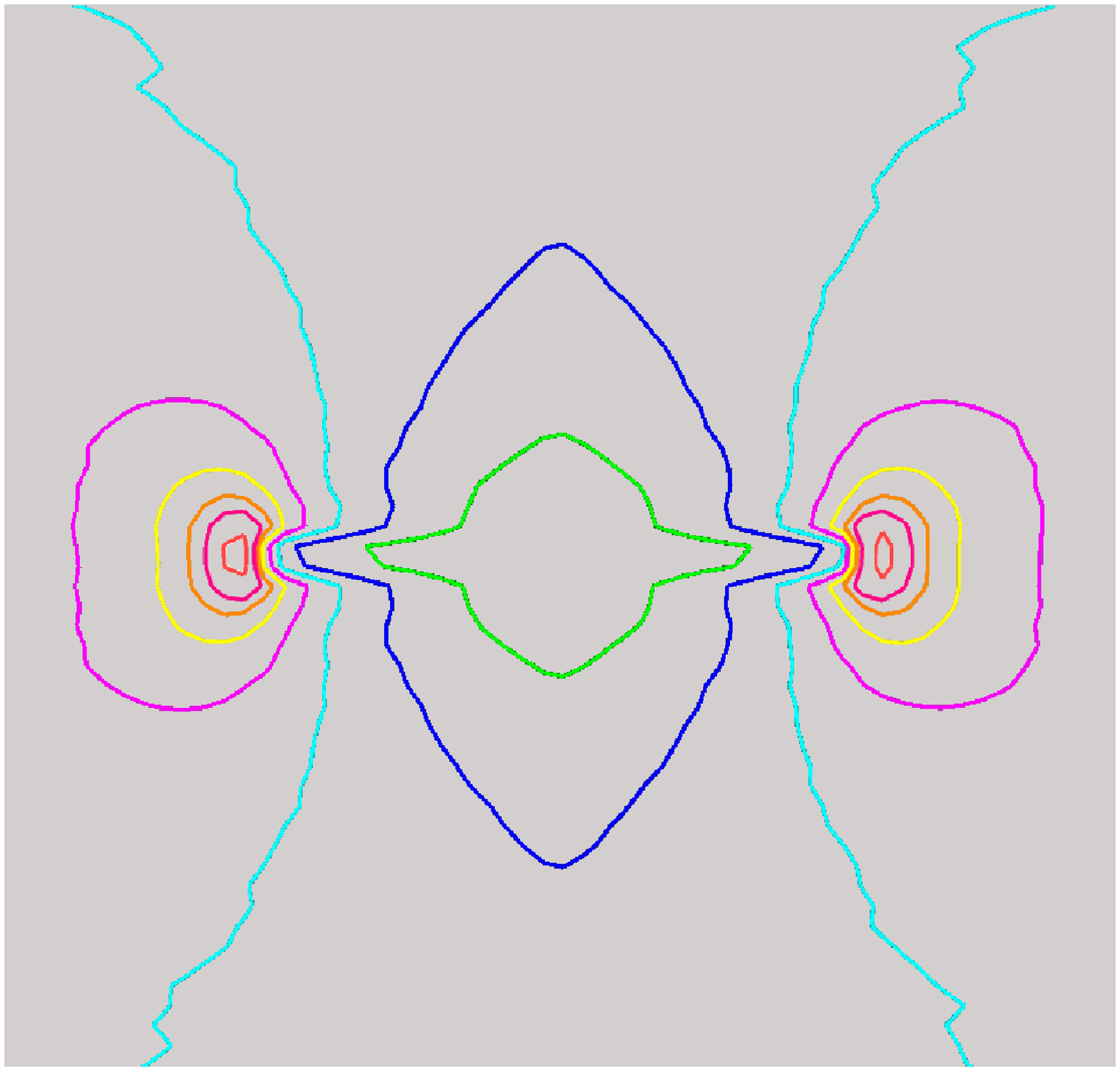}
\includegraphics[height=3.5cm]{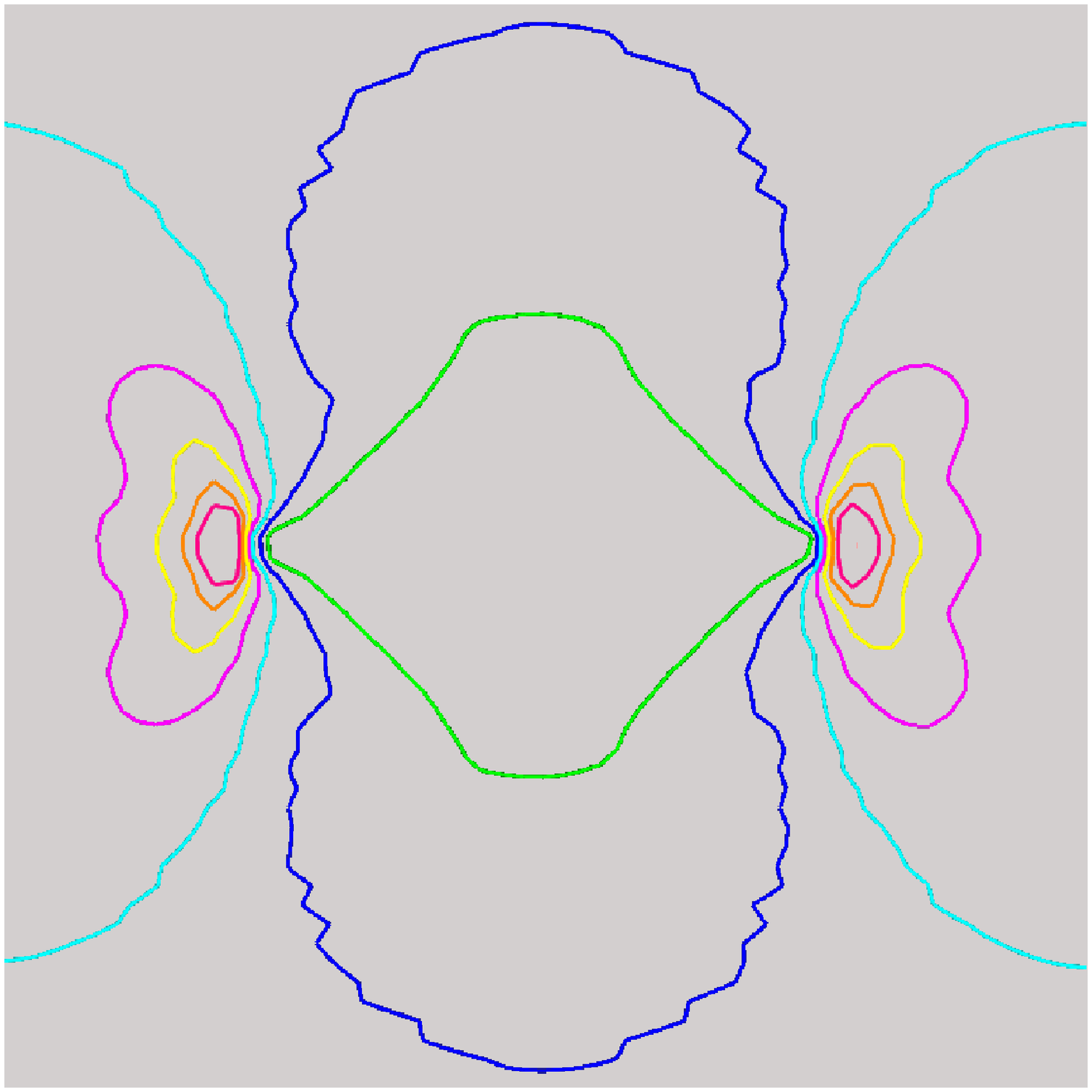}
\includegraphics[height=3.5cm]{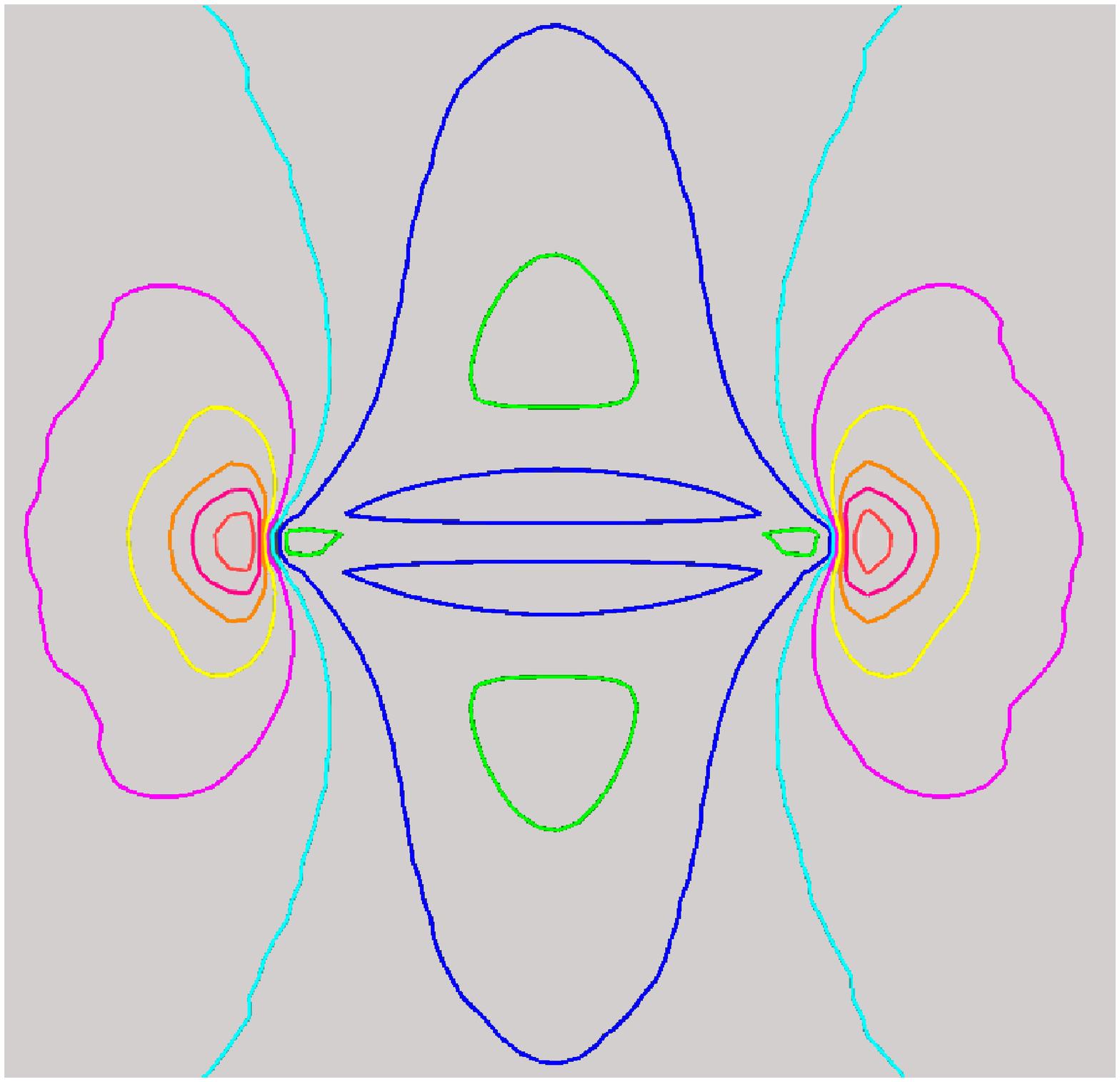}
\end{center}
\caption{The stress concentration profiles for the RFM (left), the RSM (center) and the RBM (right).}
\label{fig0}
\end{figure}

\section{Strength and size effects}

We perform numerical simulations of the models discussed above,
concentrating on the failure strength. Notice that the three model
differ mainly in the way stress is redistributed. To illustrate this 
point, we report in Fig.~\ref{fig0} the stress concentration profiles
in a triangular lattice ($L=512$) with a notch of size $a_0=16$. 
Although the angular distribution of the stress profiles differs,
the way stress decay from the crack tip is very similar, approaching
for large $L$ the $1/\sqrt{r}$ decay expected from the theory of
elasticity.

\begin{figure}
\begin{center}
\includegraphics[width=12cm]{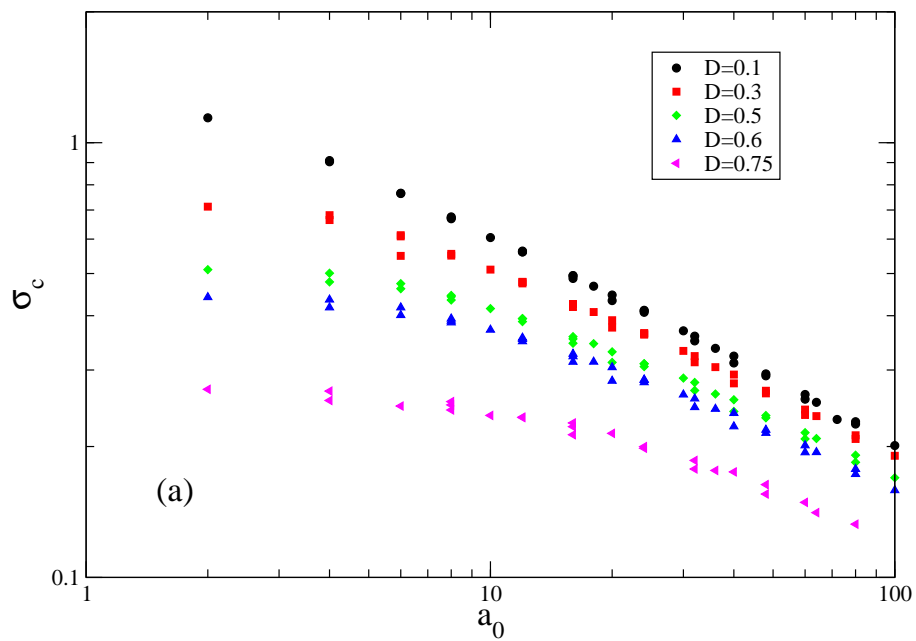}
\end{center}
\vspace{1cm}

\begin{center}
\includegraphics[width=12cm]{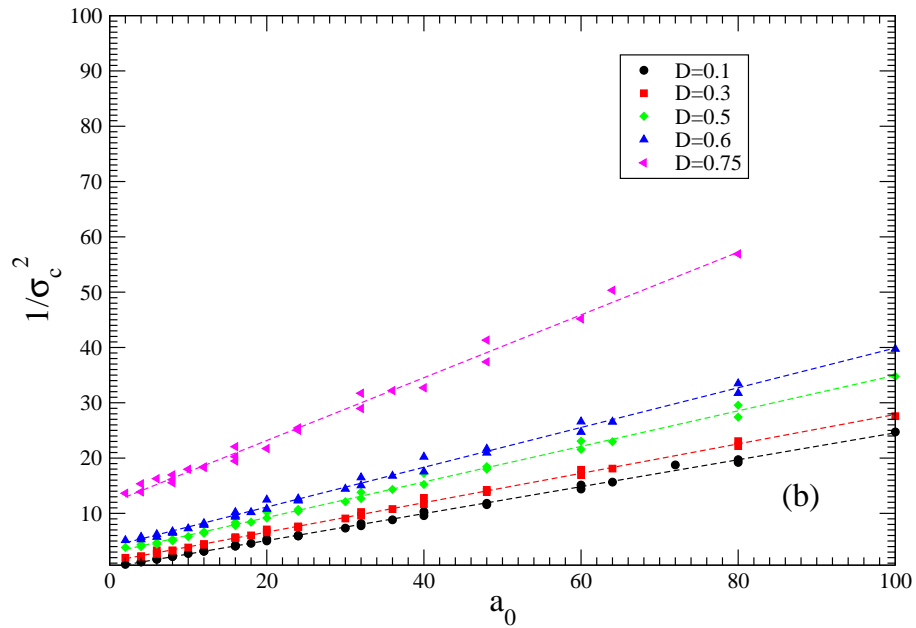}
\end{center}
\caption{a): The strength in the RFM
for several disorders $D$, notch sizes $a_0$ and
system sizes $L$. b): a scaling plot of the data according to Eq.
(\protect\ref{baz}).}
\label{fig1}
\end{figure}

\begin{figure}
\begin{center}
\includegraphics[width=12cm]{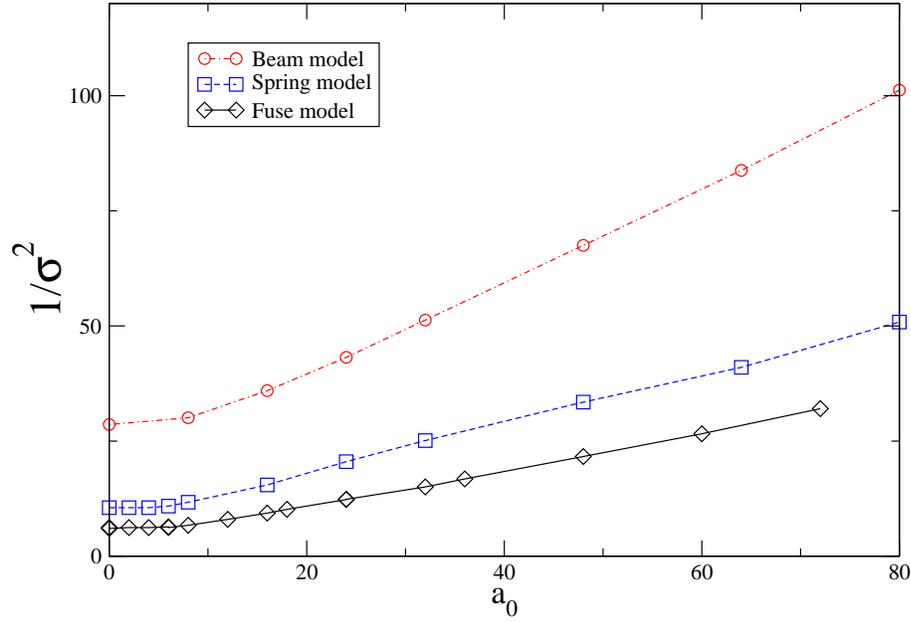}
\end{center}
\caption{ A scaling plot of the strength according to Eq.
(\protect\ref{baz}) for the RFM (for $L=192$, $D=0.6$), the RSM 
($L=256$ $D=0.5$) and the RBM ($L=256$ $D=1.0$) models.
The qualitative features of the data for different system sizes and 
disorders are the same in all the three models.}
\label{fig2}
\end{figure}

In order to obtain reliable results, the strength should be averaged over several realizations 
of the disorder. In the present simulations, we have used a minimum
of $N_r=200$ realizations and in some cases up to $N_r=8000$ realizations.
Fig. \ref{fig1}a reports the strength $\sigma_c$, averaged over different
configurations, with varying $a_0$, $D$, and $L$ for the RFM.
The most instructive way of plotting is to consider
the inverted square strength, $1/\sigma_c^2$. Assuming
Eq.~(\ref{baz}), $1/\sigma_c^2$ should become a linear
function of $a_0$ for large enough notches. Plotting the data in
this way in Fig.~\ref{fig1}b shows that for $a_0 \gg 1$, the scaling of
Eq.~(\ref{baz}) is recovered asymptotically.
Extrapolating the linear part towards $a_0 = 0$, we can define
a disorder-dependent intercept $\xi(D)$, that should be
related to the FPZ size. Furthermore, the slope of the linear part of the data
$(1/K_c^2(D))$ is also disorder-dependent, which implies a disorder-dependent
fracture toughness $G_c(D)$. Finally, a careful observation reveals that for small
$a_0$ less than a critical crack size $a_c$, the strength scaling crosses over 
from a stress concentration dominated LEFM scaling [Eq.~(\ref{baz})] to a 
disorder dominated scaling (see Fig.~\ref{fig2}). That is, for $a_0 \ll a_c$, the strength scaling 
deviates significantly from Eq.~(\ref{baz}) and saturates to a value that depends 
on disorder and the sample size, $\sigma_c (L,D)$. In particular, the 
strength of the unnotched system (for $a_0 = 0$) is finite and is smaller than the LEFM limit 
$K_c/\sqrt{\xi}$ given by Eq.~(\ref{baz}). 
In Fig.~\ref{fig2}, we present a comparison between RFM, RSM and the
RBM. The general features of the strength are the same in the three
models, indicating that only the $1/\sqrt{r}$ decay of stress
concentration is relevant for the size effect, while the precise
angular dependence of stress concentration around a notch is not
important.

In Ref.~\cite{alava08} we presented a scaling theory that
extends the earlier scaling law given by Eq.~(\ref{baz}) beyond its actual
regime of validity. A correct scaling expression has to accommodate
the three separate phenomena visibile in Fig.~\ref{fig1}: for small
notches, the dominance of statistical effects that dictate $\sigma_c
(L,D)$, the cross-over to the LEFM-like regime, and then finally an
Equation (\ref{baz}) like scaling at large $a_0$.

The cross-over takes place at a scale $a_c$ above which $\sigma_c$
follows Eq.~(\ref{baz}). For small notches, $a_0\ll a_c$ the
strength is determined by extremal statistics as is appropriate in
the limit $a_0 \rightarrow 0$ \cite{alava06}. Then one expects to
see a weak size effect, typically logarithmic in $L$. In real
materials, the scaling will depend on the damage accumulation and
the defect populations that exist in the specimens. $\sigma_c (L,D)$
is not a constant however, as Eq.~(\ref{baz})
would predict, and deviates significantly from
the LEFM-based theory, which would in general predict that 
the samples are weaker than their actual strength $\sigma_c (L,D)$.

The location of the cross-over (notch size) $a_c$ follows by
equating the strength prediction of Eq.~(\ref{baz}) and the scaling
of notchless specimes, $1/\sigma(L,D)^2 \simeq (a_c+\xi)/K_c^2$. An
appropriate scaling theory, valid for all $a_0$, is then
\begin{equation}
\frac{K_c^2}{\sigma_c^2} = \xi+a_0 f(a_c/a_0) \label{fundamental}
\end{equation}
where the statistical physics -like scaling function $f(y)$ has the
limits
\begin{eqnarray}
        f(y)\simeq
        \left\{ \begin{array}{ll}
        1  & \mbox{if \quad $y \ll 1$}  \\
        y & \mbox{if \quad $y \gg 1$}
\label{scafunc}
                \end{array}
\right. \label{scafu}
\end{eqnarray}

Thus we have for the cross-over scale
\begin{equation}
a_c \simeq (K_c(D) /\sigma_c(L,D))^2-\xi(D).
\end{equation}
For $a_0>a_c$, fracture is governed by LEFM and a scaling of the
kind of Eq.~(\ref{baz}) is recovered. The effect of disorder,
according to the scaling theory, is incorporated in the three
parameters $\sigma_c(L,D))$, $K_c(D)$,  and $\xi(D)$. In the following, we first discuss
the first two parameters and return to $\xi$ below in much more
detail. Qualitatively (since the behaviour of $\sigma_c(L,D)$ is an
independent issue entirely), the effect of
changing disorder strength for a fixed $L$ can be seen as follows. For stronger disorder, the
cross-over will take place at larger $a_c$, since the stress
concentration of the notch will be screened (as we demonstrate
below). At fixed disorder, $a_c$ increases with $L$ since notchless
specimens get weaker. The fracture toughness $G_c$ (since in the
models $E = 1$) seems in our simulations to be proportional to the
average model element strength at weak disorder at least, and
perhaps gets reduced with strong disorder \cite{alava08}. More
numerical work in this direction might be interesting.

\section{The fracture process zone}

\begin{figure}
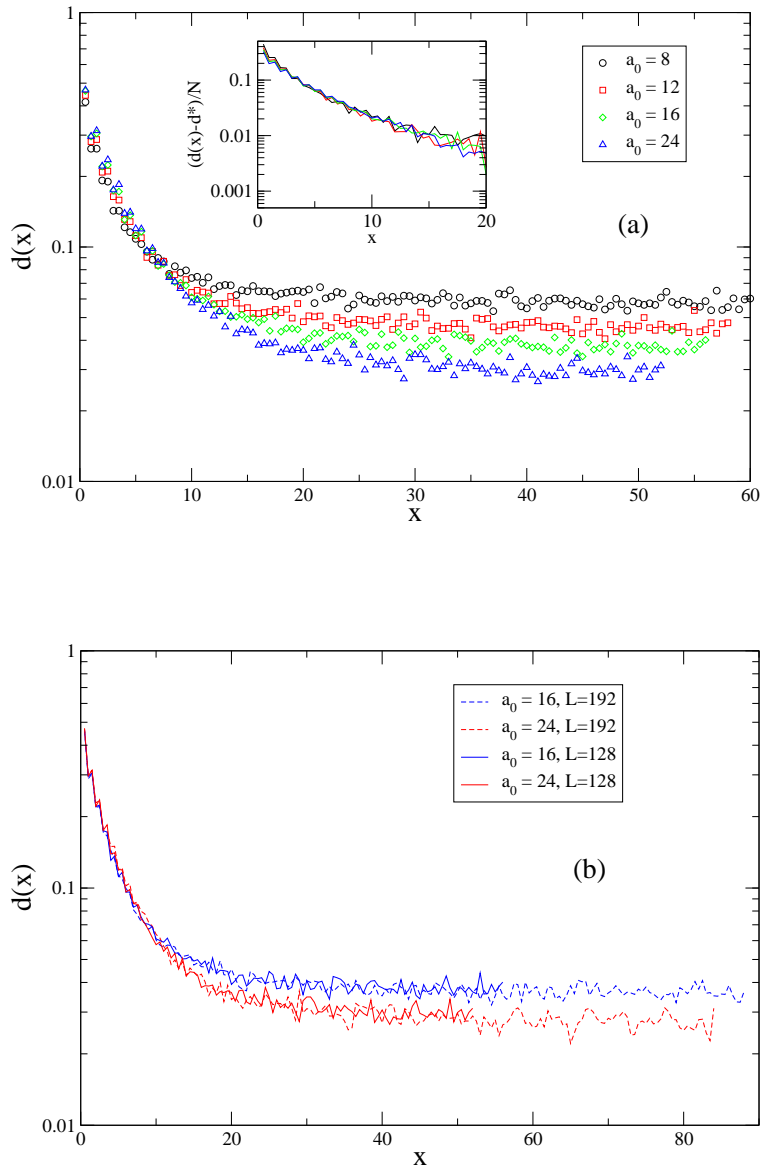

\begin{center}
\includegraphics[width=10cm]{./prof_vs_a0_a.eps}\end{center}
\vspace{1cm}
\begin{center}
\includegraphics[width=10cm]{./prof_vs_a0.eps}\end{center}
\caption{a): Damage profiles along the crack axis for
various notch sizes $a_0$. Damage profiles follow an exponential
decay on a uniform damage background, i.e.,
$d(x)= A+B\exp{(-2x/\xi_{FPZ})}$, where $A$ and $B$ are constants and
$x$ is the distance from the crack tip along the crack axis. 
In order to show that $\xi_{FPZ}$ is independent on $a_0$, we report in the
inset the profiles after subtracting the background and normalizing so that
the curves superimpose.
b) Damage profiles for two different lattice sizes $L$ and two different
notch sizes $a_0$. The profiles do not depend on $L$.}
\label{fig3}
\end{figure}

Our numerical simulations allow to monitor the damage evolution
prior to failure and can thus be used to study the development of
the FPZ. For a single realization of the disorder, we only see
diffuse damage up to the peak load, and it is difficult to 
determine the size of the FPZ. On the other hand, the FPZ can 
be clearly measured after averaging the damage
over several realizations of the disorder. Hence, the size of FPZ should thus be
considered in statistical terms as the region around the crack tip
where damage is most likely to occur. Considering for simplification
a projection of the average damage along the notch main axis
direction, we obtain a profile that is decaying exponentially
towards a homogeneous background value:
$d(x)=A+B\exp{(-2x/\xi_{FPZ})}$ (see Fig.~\ref{fig3}). The factor 
$2$ in the exponential comes from the fact that in our geometry
the FPZ extends from the two edges of the notch. We have
analyzed the data for different values of $a_0$ and $L$ in order to
check that the profiles do not depend on $L$ and on $a_0$, as long
as this is not too far from $a_c$. For $a_0 << a_c$ we naturally do
not expect to see such a "damage cloud" around the original defect.
However, it seems likely that one could measure $\xi$ around the
most critical microcrack. Recall that in this regime one expects the
strength to saturate at $\sigma_c = \sqrt{K_c/(a_c+\xi)}$.

\begin{figure}
\begin{center}
\includegraphics[width=12cm]{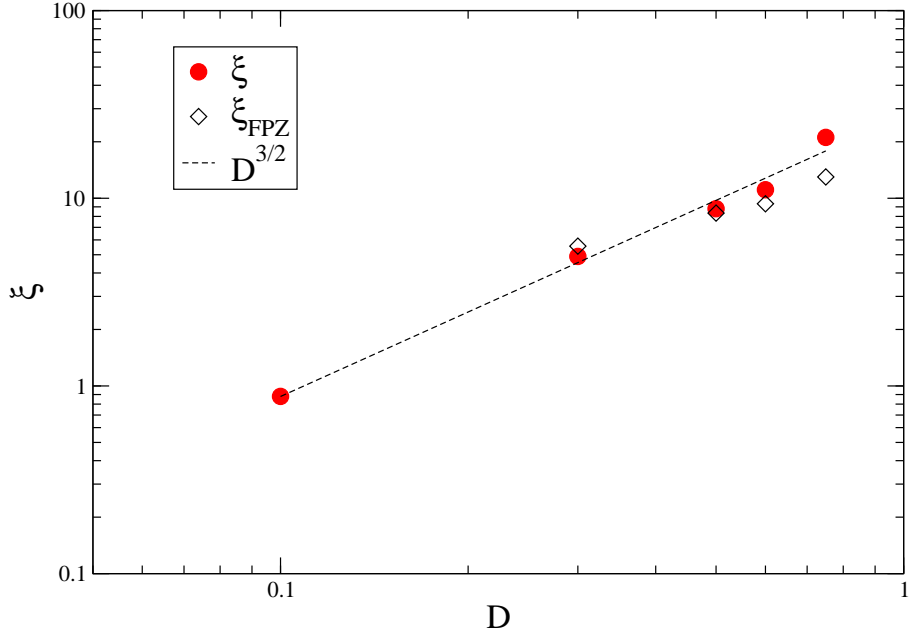}\end{center}
\caption{The dependence of the FPZ size on disorder for the RFM ($L=128$, $a_0=16$). 
Estimates from the strength ($\xi$) and from damage profiles $\xi_{FPZ}$ are similar.
We could not obtain reliable estimates from damage profiles for very weak 
disorder ($D=0.1$).}
\label{fig4}
\end{figure}

\begin{figure}
\begin{center}
\includegraphics[width=12cm]{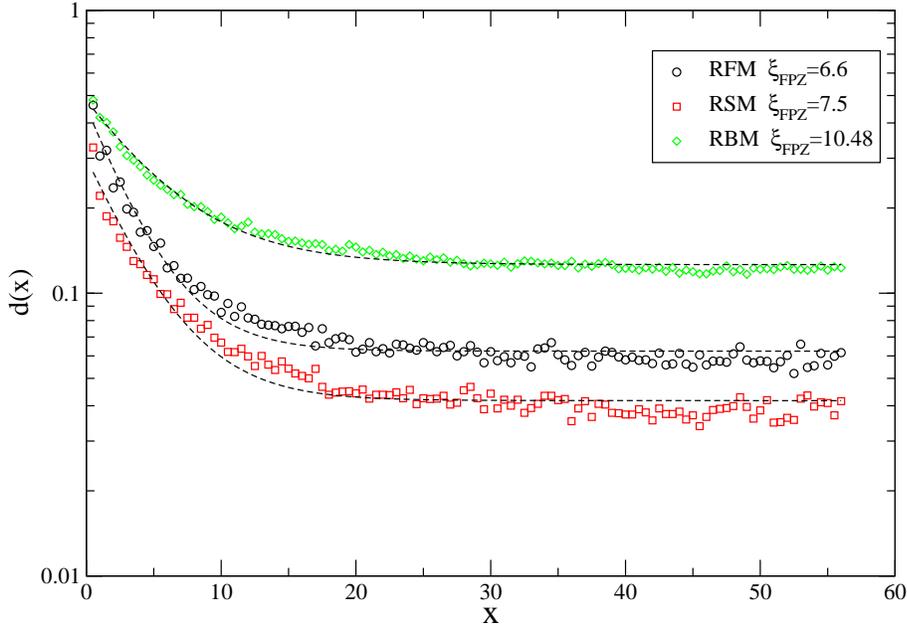}\end{center}
\caption{A comparison of the damage profiles measured in RFM, RSM and RBM using 
system size $L = 128$, disorder $D = 0.6$, and an initial notch size $a_0 = 16$. 
Two thousand samples are used for averaging the damage profiles. The result show
that while the profiles are qualitatively similar $\xi_{FPZ}$ and the damage
backgound differ.}
\label{fig5}
\end{figure}

Notice that the LEFM stress intensity factor would indicate a
$1/\sqrt{r}$ -like divergence of the stress at the crack tip. It is
evident that the observed exponential decay of damage profile $d$ is
in contrast to a $1/\sqrt{r}$ -like decay and should be naturally
interpreted as a screening of the crack tip caused by the disorder.
In fact, the FPZ size $\xi_{FPZ}$ depends on the disorder strength
$D$ as shown in Fig.~\ref{fig4}. The data can be roughly described
by a power law relation $\xi_{FPZ} \sim D^{3/2}$. As discussed in
Ref.~\cite{alava08}, if we plot the fracture process size
$\xi_{FPZ}(D)$ against the intrinsic scale $\xi$ resulting from the
fits of the strength data to Eq.~(\ref{baz}), we obtain a linear
relation. Hence, we can conclude that $\xi$ is indeed a direct
measure of the FPZ size. Fig.~\ref{fig5} reports a comparison
of the damage profiles in RFM, RSM and RBM. It can be seen that
the results are qualitatively similar for all the three models considered. The value
of the FPZ size $\xi_{FPZ}$, however, differs slightly for the 
for the three models.

\begin{figure}
\begin{center}
\includegraphics[width=12cm]{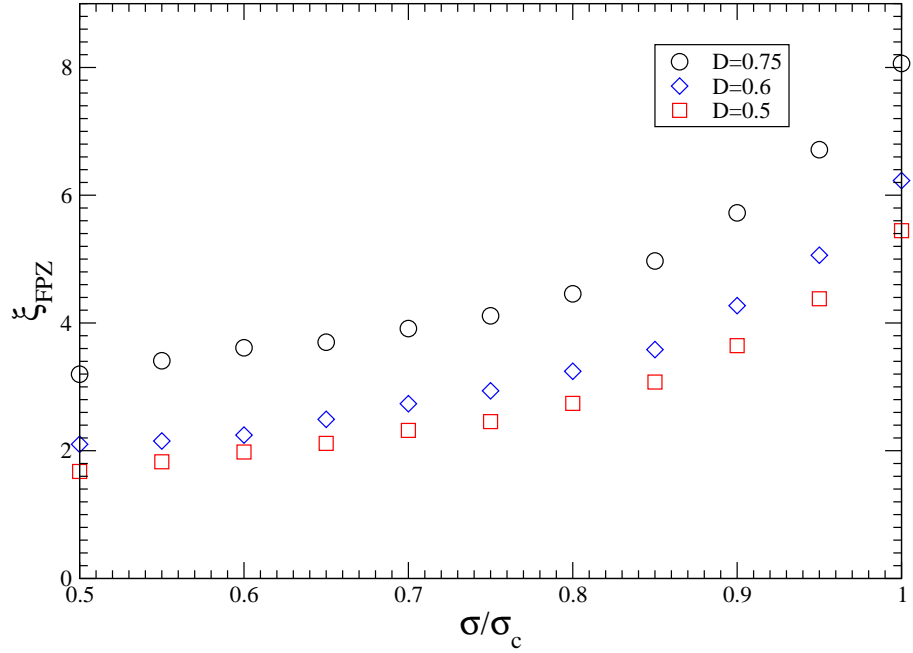}\end{center}
\caption{The FPZ size as a function of the applied stress normalized by the peak stress
as obtained from damage profiles.
Data are for RFM for different values of disorder ($a_0=16$, $L=128$).}
\label{fig6}
\end{figure}

\begin{figure}
\begin{center}
\includegraphics[width=12cm]{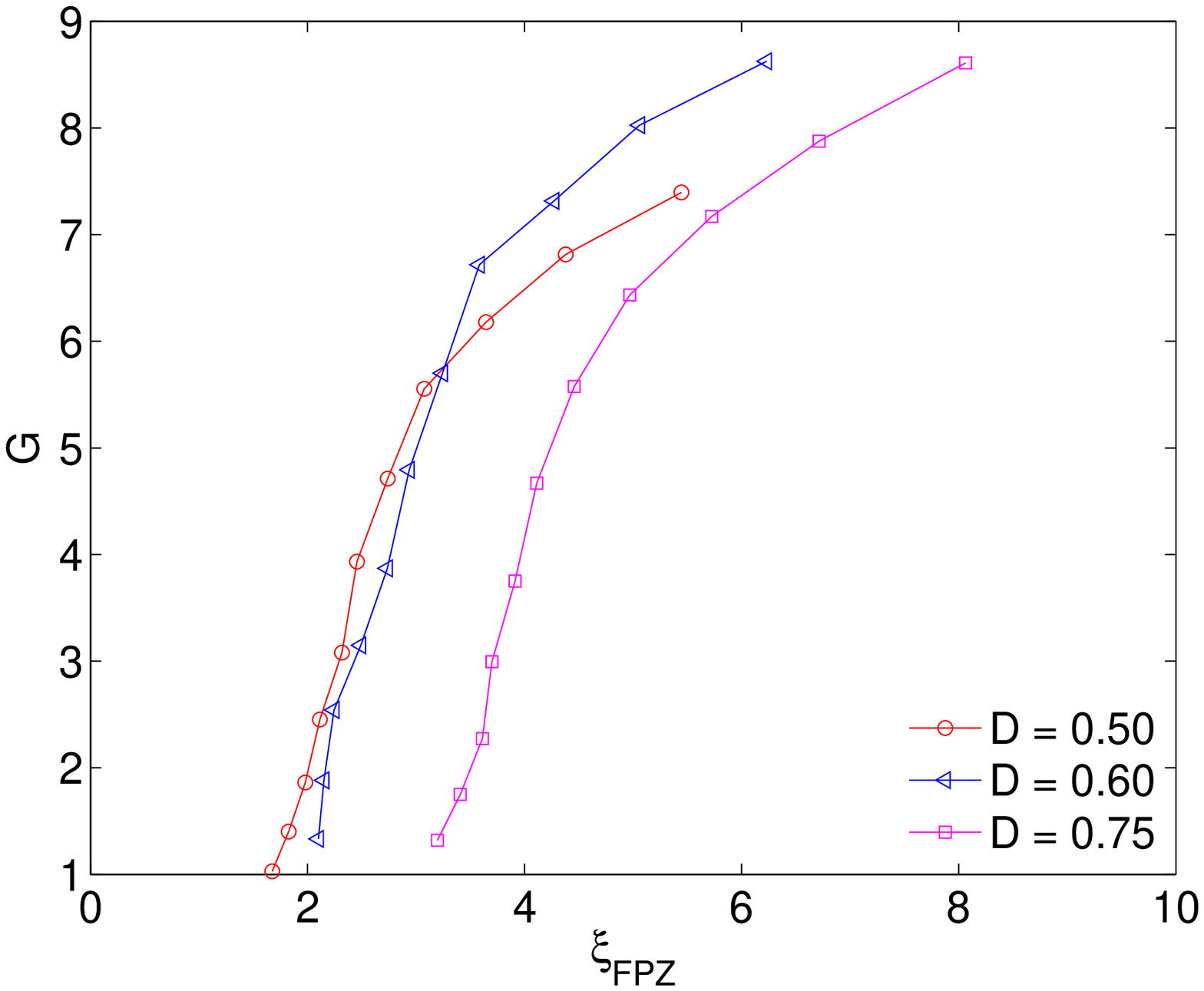}\end{center}
\caption{The R curve of the RFM for different values of the disorder D.}
\label{fig7}
\end{figure}

The FPZ progressively develops before the peak load by damage
accumulation. To visualize this process, we have computed damage
profiles at different values of the applied stress. One can then
obtain the FPZ size $\xi_{FPZ}(D)$ as a function of the stress. As
can be seen in Fig.~\ref{fig6}, there is a gradual increase of
$\xi_{FPZ}(D)$ with stress. This growth relates to the R-curve of
the material \cite{morel00} which is usually defined in terms 
of the elastic energy released due to crack growth 
$G \equiv {\partial U}/ {\partial a}$ \cite{bazantbook}. 
In the RFM model, we can derive $G$ from the lattice ``elastic'' energy $U=I^2/(2\Sigma)$, 
where $\Sigma$ is the conductivity, as
\begin{equation}
G\approx  \frac{\Delta U}{\Delta a} = \frac{I^2}{2\Sigma^2}\frac{\Delta\Sigma}{\Delta a}  
\end{equation}
where $\Delta\Sigma = (\Sigma_0 - \Sigma)$ is the conductivity change after the crack has
extended by $\Delta a$ such that $a = a_0 + \Delta a$, 
and $\Sigma_0$ is the initial conductivity. We report the R-curve for different values of
disorder in Fig.~\ref{fig7}. The data is shown for the current values 
$I \in [I_c /2,]$ in which  $\xi(I)$ can be extracted from the 
damage profiles with a reliable accuracy. In this regime the R-curves show in
general two behaviors: an initial rapid increase due to the accumulating damage
that changes the average conductivity, followed by a saturation as the FPZ starts to increase even more rapidly. The conductivity change is a mean-field phenomenon that accounts for the total damage in the system, proportional to $D(I)$, and thus to the failure thresholds $P(i_c)$. The growth of  $\xi(I)$ is not expected to be so simply related to $D$. It is interesting to note that as a result the R-curves for various disorders overlap in the manner depicted in Fig.~\ref{fig7}. The size-dependence of these R-curves would be expected to be negligible as long as the strength is governed by Eq. (\ref{baz}).

\section{Conclusion}

We have resorted to simulations of statistical fracture models to
analyze the problem of the size-effect in the failure of
quasi-brittle materials. For large notches, the simulations recover
the expression based on LEFM \cite{bazant99,bazant00,bazant04b} and
allow to relate the effective FPZ size $\xi$ to the actual average
damage profiles. As the notch size is decreased we observe a
crossover at a novel scale $a_c$ to a disorder-dominated
size-dependent regime that is not described by LEFM and is
furthermore seen in experiments \cite{alava08}. All the regimes are
summarized in a generalized scaling expression for the strength of
disordered media.

Several interesting future questions remain, like theoretical
computations of parameters such as $a_c$, $K_c$, and the detailed
understanding of the origin and shape of the statistical FPZ.  These
would be in particular important in order to help to achieve
practical predictions. Recall our results have shown, that all such
parameters are dependent on disorder, which in the models used
translates into the damage accumulated at a given local stress. This
quantity is evidently hard to access experimentally, but is
unfortunately theoretically necessary. For such reasons, it would be
relevant to investigate three dimensional systems and possibly look
at other kinds of disorder (eg. locally varying elastic moduli).

{\bf Acknowledgments -} MJA would like to acknowledge the
support of the Center of Excellence -program of the Academy of Finland.
MJA and SZ gratefully  thank the financial support of the European Commissions
NEST Pathfinder programme TRIGS under contract NEST-2005-PATH-COM-043386.
PKKVN acknowledges support from Mathematical, Information and
Computational Sciences Division, Office of Advanced Scientific
Computing Research, U.S. Department of Energy
under contract number DE-AC05-00OR22725 with UT-Battelle, LLC.
PKKVN also acknowledges the use of IBM BG/L
resources made available to him at Argonne National
Laboratory through INCITE.


\end{document}